# Stimulated Raman Spectroscopy with Tunable Visible Broadband Probe Pulse Generated by Kerr-Instability Amplification

Nathan G. Drouillard[1*], TJ Hammond[1]

[1]Department of Physics, University of Windsor, 401 Sunset Ave., N9B 3P4, Windsor, Ontario, Canada

*Corresponding author email: droui116@uwindsor.ca

ORCiDs

Nathan Drouillard: 0000-0002-6180-7321

TJ Hammond: 0000-0001-8334-4332

**Abstract**

Femtosecond, broadband stimulated Raman spectroscopy is a popular approach to measuring molecular dynamics with excellent signal-to-noise and spectral resolution. We present a new method for broadband stimulated Raman spectroscopy that employs Kerr-instability amplification to amplify the supercontinuum spectrum from sapphire and create a highly tunable Raman probe spectrum spanning from 530 to 1000 nm (-6000 to 2800 cm$^{-1}$). Our method, called Kerr-instability amplification for broadband stimulated Raman spectroscopy (KAB-SRS) provides an alternative to optical parametric amplifiers by producing a broader and more tunable spectrum at a significantly reduced cost to OPA implementations. We demonstrate the effectiveness of KAB-SRS by measuring the stimulated Raman loss spectrum of 1-decanol.

**Graphical abstract**

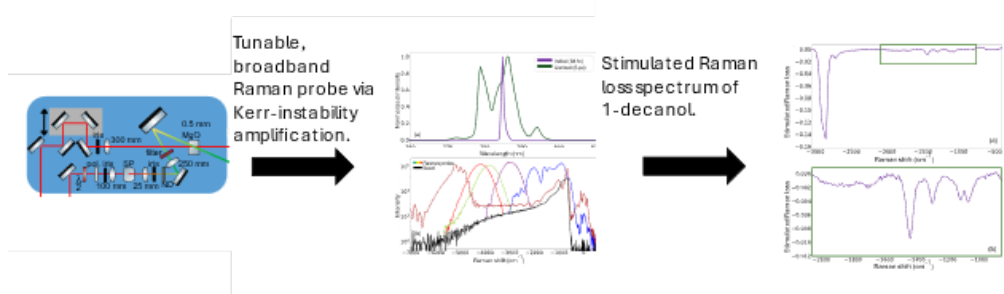

**Introduction**

With molecular vibrations occurring on the timescale of femtoseconds (1 fs = $1 \times 10^{-15}$ s) and rotations for small molecules on the order of picoseconds (1 ps = $1 \times 10^{-12}$ s), spectroscopic measurements that probe the structure of matter benefit greatly from ultrafast light sources.[1] While there exist ultrafast electronic spectroscopies, these techniques often make it difficult to measure structural dynamics.[1,2] Therefore, vibrational spectroscopy is a more suitable approach to elucidating the desired molecular information.

Most molecules possess several Raman active modes that provide information on molecular vibrations, making Raman spectroscopy a straightforward alternative to infrared (IR) spectroscopy, which requires far-IR light sources and detectors. Especially with regards to aqueous samples that are dominated by infrared absorption, Raman spectroscopy is a viable alternative.[3,4]

Femtosecond stimulated Raman spectroscopy (FSRS) has become a common approach to studying molecules and reaction pathways on ultrafast timescales.[5-7] FSRS offers several advantages over spontaneous Raman, namely its fast acquisition time, superior signal-to-noise ratio, and fluorescence rejection making it a background-free technique.[2,8] The theory of FSRS has been published in detail elsewhere.[1,2]

Typically performed as a three-pulse experiment, FSRS can be thought of as stimulated Raman spectroscopy (SRS) with an additional pulse called the actinic pump pulse that provides the temporal resolution of the measurement. Without the actinic pump, but while still having a femtosecond, broadband Raman probe pulse, the name broadband stimulated Raman spectroscopy (BSRS) is often used.[9,10] In this work, we are not concerned about the temporal resolution provided by an actinic pulse, but we sought to develop a novel technique for generating the broadband Raman probe pulse, which ultimately defines the bandwidth of detection in BSRS. Our efforts are aligned with recent developments in the field of FSRS which have been focussed on the tunability of ultrafast laser pulses.[11, 12]

A common approach to generating a broadband probe spectrum is supercontinuum generation in sapphire.[2] Downsides to supercontinuum generation include intense residual pump light and less intense light at longer wavenumbers. Alternatively, one may use an optical parametric amplifier (OPA) or its noncollinear counterpart (NOPA).[2,13,14] The bandwidth of this method is typically limited to 1200 cm$^{-1}$ due to gain narrowing[2,15] and is limited to gain media that possess $\chi^{(2)}$ optical nonlinearities, far less common than those that exhibit a $\chi^{(3)}$ response.[16] Fang et al have developed an innovative alternative that uses cascaded four-wave mixing (CFWM) to produce multiple sidebands, each having a FWHM bandwidth of up to 1600 cm$^{-1}$, with a total spectral region of detection spanning 100-4000 cm$^{-1}$.[17,18] Yet still, one must scan using multiple Raman probe beams to span the entire Raman spectrum from 0-4000 cm$^{-1}$. Conversely, we present a new method that generates a tunable, broadband probe that spans from -6000 cm$^{-1}$ to 0 cm$^{-1}$ on the anti-Stokes side, making it possible to make such measurements with one individual probe spectrum. Furthermore, we can generate a probe spectrum that spans from 0 to 2800 cm$^{-1}$ on the Stokes side, by changing the filter used prior to amplifying the supercontinuum.

Since we operate with a blue-shifted probe in this work, we detect stimulated inverse Raman scattering (IRS), [19-21] sometimes referred to as Raman loss. Hence, techniques such as ours are often referred to as stimulated inverse Raman spectroscopy, anti-Stokes (femtosecond) stimulated Raman spectroscopy, or ultrafast Raman loss spectroscopy (URLS).[22-25] For this work, we will use the convention that IRS refers to the physical process that is measured via URLS.[25] Stimulated inverse Raman scattering is an intensity dependent process where the intensity of the anti-Stokes field grows exponentially with the pump intensity, $I_{as}(z) \propto \exp(|g|I_{pu}z)$, with $g = \frac{2\pi\omega_{pu}}{nc} \text{Im}[\chi_{NL}]$ and where $\chi_{NL}$ is the nonlinear susceptibility.[21,26]

We present Kerr-instability amplification for broadband stimulated Raman spectroscopy (KAB-SRS), which uses Kerr-instability amplification (KIA) rather than CFWM. KIA extends four-wave mixing to the high-intensity regime, leading to a relaxed phase-matching condition and enabling broadband amplification across nearly an octave of bandwidth. The amplified spectrum from KIA is tunable to both the blue (anti-Stokes) and red (Stokes) sides of the pump wavelength. The theory for KIA and the realization of broadband amplification are outlined in our previous work.[27-30] As established in our prior research, KIA provides a cost-effective, versatile route for generating broadband Raman probe spectra. KIA can amplify down to 500 nm with a 785 nm pump (-7000 cm$^{-1}$) in one single spectrum, but we are limited by dispersion of the reimaging optics that come before amplification rather than phase matching in the case of an OPA or NOPA.[25] We show the efficacy of our method by measuring the Raman loss spectrum of 1-decanol.

**Methodology**

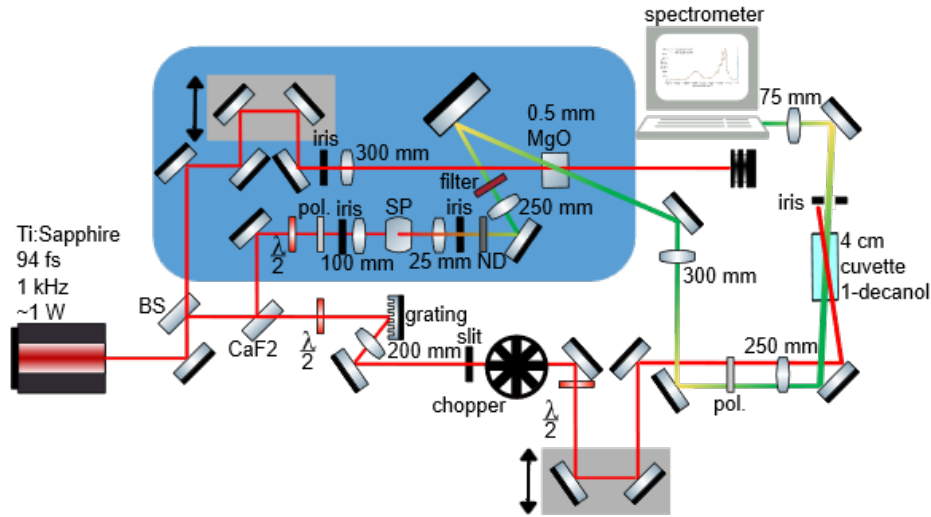

Figure 1: Experimental setup. We split the output of a Ti: Sapphire laser for Kerr-instability amplification (KIA) and generating the Raman pump pulse. The blue shaded region is the setup for KIA, where the KIA pump is overlapped temporally and spatially with the supercontinuum seed, produced by a sapphire plate (SP). We use KIA to amplify the supercontinuum and create the broadband Raman probe. The relative temporal delay of the KIA pump and seed is tuned to amplify the desired probe wavelength. The Raman pump pulse is created by limiting the laser spectrum via a grating and an adjustable slit. The Raman pump is overlapped with the Raman probe in the 4 cm sample cell which contains the 1-decanol sample. We measure the modulated Raman probe spectrum with a spectrometer.

We use a 1 W, 1 kHz Ti: Sapphire laser (Quantronix IntegraC) with a pulse duration of about 94 fs and a central wavelength of 785 nm. The beam is first split by a beamsplitter (BS). The transmitted part is the KIA pump, while the reflected part is split again by a $CaF_2$ window. The reflection from the $CaF_2$ is the seed for KIA, and the transmission becomes the Raman pump.

For KIA, the KIA pump passes through a delay stage to achieve temporal overlap with the seed. The power of the KIA pump is controlled to 180 mW by an iris that precedes the 300 mm lens which focuses the beam to a waist $w_0$ of 70 μm at the 0.5 mm MgO sample. This gives a peak intensity of approximately $1.4 \times 10^{17} W/m^2$ for the KIA pump, just below the damage threshold for MgO.[31] The MgO sample is oriented in the 110 plane to maximize amplification.[32]

The power of the seed for the KIA is controlled by a wire grid polarizer and halfwave plate (λ/2) to be 1 mW before the 5 mm thick sapphire plate. The result is a stable supercontinuum spectrum that is amplified by the KIA pump. A 750 nm short pass filter is placed prior to the amplification process to optimize the amplification at the desired wavelength and attenuate the 785 nm part of the spectrum. The supercontinuum seed pulse is focused with a 250 mm lens and guided by a 50 mm diameter mirror to the MgO sample, where it is spatially overlapped with the KIA pump. The angle between the KIA pump and the seed is set to 5.1 degrees, enabling supercontinuum amplification,[27] and the temporal delay is tuned to maximize the 725 nm part of the amplified spectrum, which serves as the Raman probe.

We operate the KIA setup below saturation, not in the maximal amplification regime. This choice is made to reduce saturation effects in the amplification process that can lead to an unstable spectral amplitude, and to limit the amount of 785 nm light that is amplified. We use the variable neutral density (ND) filter to attenuate the seed prior to amplification and produce a stable Raman probe spectrum. While it is possible to amplify the power of the supercontinuum by 4 orders of magnitude, we choose to operate in a regime where the power of the probe at 635 nm compared to the unamplified supercontinuum is greater, on average, by two-three orders of magnitude.

The Raman pump pulse is created by spectrally limiting the original laser. The combination of a grating (Thorlabs GR25-1208, 1200 lines/mm) and an adjustable slit (Thorlabs VA100/M) significantly narrows the spectrum. Fig. 2(a) shows the laser spectrum in green and the limited spectrum (which serves as the Raman pump) in purple. The bandwidth of the laser spectrum is 10 nm, corresponding to a transform-limited pulse duration of 94 fs, which is consistent with our previously measured pulse duration of approximately 100 fs using a home-built frequency-resolved optical gating (FROG) setup. We calculate a Raman pump bandwidth of about 1 nm, which approaches the resolution of our spectrometer. We perform an intensity autocorrelation of the pulse and measure a full width at half-maximum (FWHM) pulse duration of 6 ps.

The grating efficiency is optimized by rotating the polarization from vertical to horizontal polarization before the grating using a halfwave plate. A narrower Raman spectrum is advantageous due to improved spectral resolution. The Raman pump is modulated by a chopper (Thorlabs MC1F10HP) that runs at 100 Hz. The laser TTL signal of 1 kHz is sent to the chopper controller (Thorlabs MC2000B), the 1 kHz signal is converted to 100 Hz to control the chopper. The 100 Hz output signal of the chopper is sent to an Arduino Uno, where it is converted to a 200 Hz signal, to account for the 500 μs delay that it takes for the spectrometer to read an input trigger signal. The Arduino output is connected to the spectrometer (Ocean Optics Flame-S), which has an integration time of 2 ms, allowing it to measure 2 pulses "on" followed by 2 pulses "off".

Since the Raman pump is horizontally polarized, it is rotated back to vertical polarization by a second halfwave plate. The Raman pump and probe are then passed through the same wire grid polarizer to ensure they have the same polarization. Next, the beams are passed in parallel through a 250 mm lens, focusing the pulses together within the sample. The Raman pump and

Raman probe have peak intensities of approximately $6.0 \times 10^{13} W/m^2$ and $4.0 \times 10^{14} W/m^2$, respectively, at the focus. The probe beam waist ($w_{0_{pr}} \approx 30$ μm) fits within the pump beam waist ($w_{0_{pu}} \approx 350$ μm). We use a smaller probe beam waist relative to the pump to maximize our signal.[33] Consequently, the probe beam is more intense than the pump, but this is acceptable because the Raman loss signal scales linearly with probe intensity.[21,22]

If we operate our KIA setup in the maximum amplification regime of three-four orders of magnitude, we generate probe intensities on the order of $10^{15} - 10^{16} W/m^2$. At such intensities we observe additional nonlinear effects in the 1-decanol sample, such as four-wave mixing and cross-phase modulation, which interfere with the Raman spectrum.[34] Further investigation of the optimal pump and probe intensities is needed.

The pure 1-decanol sample is contained in a glass sample cell with an optical path length of 4.0 cm, centered at the foci. The two beams diverge after the sample, the Raman probe passes through an iris that blocks the pump, and the probe spectrum is measured by the Ocean Optics Flame-S spectrometer. We have previously shown that the pulse duration of the probe is approximately 42 fs,[27] and we have previously measured 64.6 fs$^2$/mm for the group velocity dispersion (GVD).[35] Given a propagation distance of 4 cm, we calculate that the pulse duration will increase to 176 fs by propagating through the 4 cm sample cell, and this pulse duration is still much shorter than the Raman pump. We estimate the spatial overlap of the Raman pump and probe to be 2.5 cm, which is further substantiated by our observation that the signal is optimized from 2-4 cm long samples. Further discussion on pump-probe overlap will be discussed below.

Our method provides a clear advantage over spontaneous Raman spectroscopy in terms of acquisition time. We have previously measured the spontaneous Raman spectrum of 1-decanol using a 50 mW, 532 nm pen laser in a simple geometry and an integration time of 2s.[36] We have also measured the spontaneous Raman spectrum using a Raman microscope, where we employed a 532 nm CW laser with an average power of 32 mW and integrated for 0.1ms 20 times. While it is possible to measure 1-decanol using either of these spontaneous Raman methods, this work is meant to illustrate a new way of creating a broadly tunable Raman probe spectrum, enabling future developments of FSRS experiments that might seek to use a variety of Raman pump wavelengths to match a resonance with a particular molecule. Although the strong $-2900$ cm$^{-1}$ mode is visible within just four spectra, we acquire spectra for 2 minutes to resolve the weaker modes near $-1500$ cm$^{-1}$ and improve the signal-to-noise ratio, corresponding to a total of 60,000 spectra.

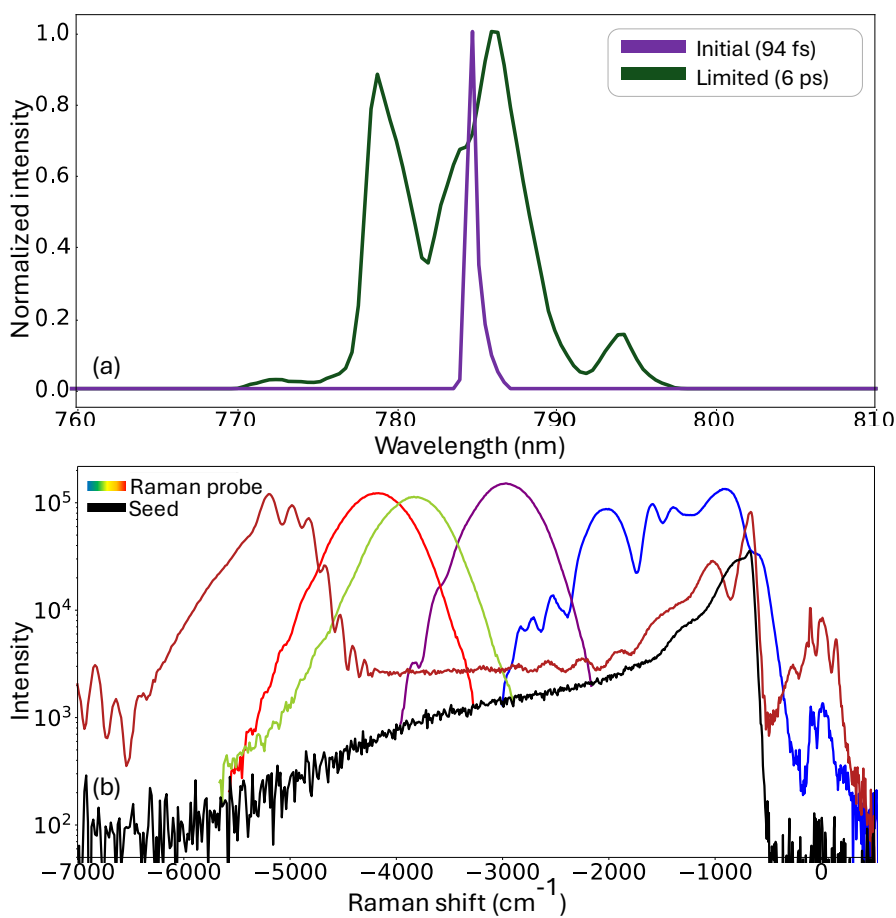

Figure 2: (a) KIA pump (green) and Raman pump (purple) spectra. The legend indicates the full width at half-maximum (FWHM) pulse durations. (b) adjusting the delay between the KIA pump and seed (black) allows the amplified broadband probe spectrum (various colours denoting various delays) to be tuned from -6000 cm$^{-1}$ to 0 cm$^{-1}$. The use of the 750 nm (-595 cm$^{-1}$) cutoff filter reduces amplification near 0 cm$^{-1}$ in this case, but amplification is otherwise possible at lower energies.

The strength of our method becomes clear when considering Fig. 2 (b), which illustrates not only the level of tunability, but also the bandwidth that is available for the Raman probe. The seed spectrum is shown by the black line, and different delays are indicated by different colours in the plot. By adjusting the delay between the KIA pump and seed, the Raman probe spectrum can span from nearly -6000 to 0 cm$^{-1}$ (530 nm to 785 nm). We note once again that we are using a 750 nm (-595 cm$^{-1}$) cutoff filter prior to amplification. Without a filter, we can amplify from 533 to 1000 nm (approximately -6000 to 2800 cm$^{-1}$), previously not possible in a single crystal. When a high-frequency seed pulse is amplified at around -5500 cm$^{-1}$ (530 nm), we observe a peak amplification of $10^3$ and a factor of 2 amplification that extends across the supercontinuum. Given a Raman probe pulse centered at 530 nm with a bandwidth of 2000 cm$^{-1}$, it is possible with our method to use a short wavelength Raman pump pulse, which is advantageous for experiments that employ a pre-resonance enhancement.[18,37] The 2000 cm$^{-1}$ bandwidth is not limited by the amplification process, but rather by the dispersion of the lenses used between the sapphire and MgO.

**Results**

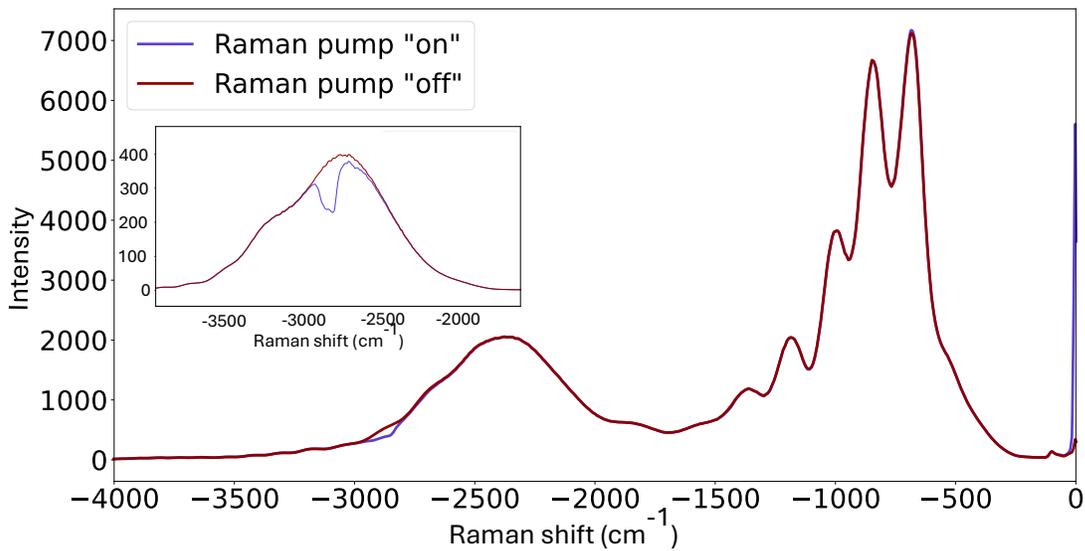

Figure 3: Raman probe spectrum with Raman pump off is shown in maroon. With the Raman pump present, stimulated inverse Raman scattering occurs, indicated by the dip in the lavender

line near -2900 cm$^{-1}$. The inset shows the probe spectrum when maximized at -2900 cm$^{-1}$, showing the feature more clearly.

The supercontinuum (seed) spectrum resulting from sapphire is amplified by the KIA pump and optimized to span from -3000 to -500 cm$^{-1}$ to detect the strongest Raman modes of 1-decanol. When the Raman pump pulse is included (lavender), there is a clear inverse Raman feature near -2900 cm$^{-1}$.

While the stimulated Raman signal is detectable with the supercontinuum (seed) spectrum from sapphire, the chirp of the spectrum is not favourable for detecting weak signals. The chirp of a supercontinuum spectrum can be compensated for with a prism pair[8]. By amplifying the seed spectrum via KIA, we produce a smoother Gaussian spectrum at the desired wavenumber, improving the utility of the probe spectrum. Furthermore, amplifying the desired probe wavelength improves the signal-to-noise ratio of the measured stimulated Raman signal by a factor of 17 in our experiments.

By amplifying the supercontinuum spectrum, we generate an ultrashort, widely tunable probe spectrum. We have previously shown that it is possible to compensate for the chirp of the pulse prior to amplification, resulting in a near transform-limited pulse[30]. Compressing the probe spectrum will be the subject of future work, which may facilitate detecting weaker signals.

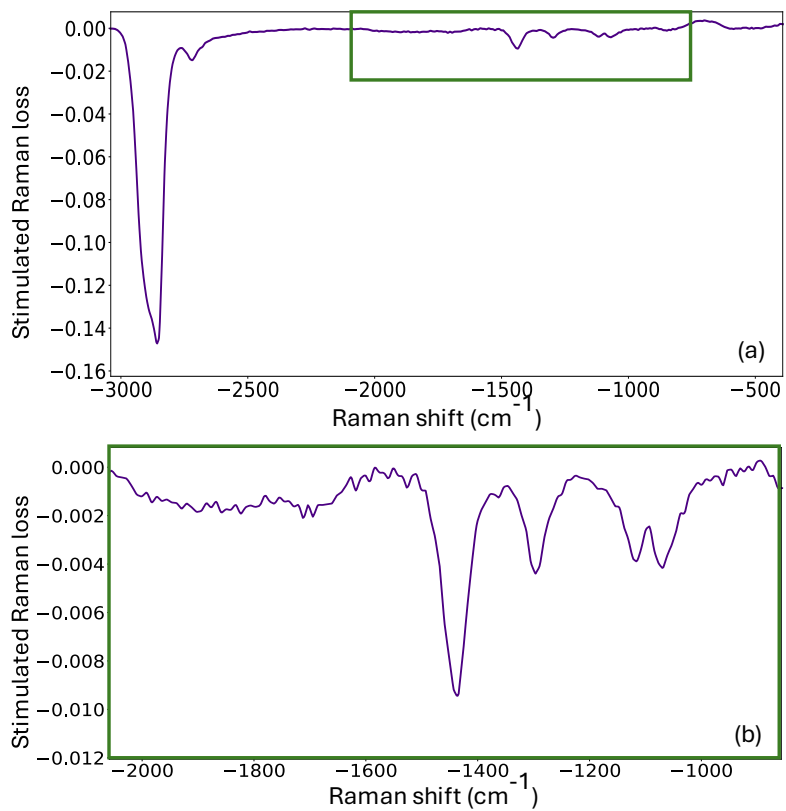

Figure 4: (a) Stimulated Raman loss spectrum calculated by dividing the lavender curve by the maroon curve in Fig. 3. Shown are the overtone mode at -2858 cm$^{-1}$, the CH$_2$ and CH$_3$ stretching modes ranging from 2880 to 2957 cm$^{-1}$, and the combination mode at -2720 cm$^{-1}$. (b) tuning the delay to maximize the probe spectrum at -1500 cm$^{-1}$, we can resolve additional known Raman modes at lower energies.

The stimulated Raman loss spectra shown in Fig. 4 are obtained from applying the well-known equation Raman gain (loss) = $\log_{10}(\frac{\text{Pump on}}{\text{Pump off}})$ to the data in Fig. 3, where the measured probe spectrum while the Raman pump is "on" is divided by the spectrum with the pump "off"[2]. Visually, this operation becomes obvious upon re-examining the lavender ("on") and maroon ("off") lines of Fig. 3. Figure 4(a) shows the full Raman loss spectrum that we measure with the probe spectrum shown previously in Fig. 3. The strongest peak near -2900 cm$^{-1}$ is associated with the CH$_2$ and CH$_3$ stretching and overtone modes of 1-decanol, ranging from -2880 to -2957 cm$^{-1}$.[39] The peak of the inverse Raman signal occurs at -2858 cm$^{-1}$, corresponding to the overtone mode.[39] Therefore, it is likely that we are exciting the molecule to the second excited state and driving the methyl stretching modes. In addition, the shoulder of the main peak at -2720 cm$^{-1}$ corresponds to a known combination mode.[39]

Although 1-decanol has other Raman active modes at lower energies ranging from -1500 to -1000 cm$^{-1}$, they are much weaker than those near -2900 cm$^{-1}$. Coupled with the chirp of our probe spectrum from -2000 to -1000 cm$^{-1}$, characteristic of both the sapphire supercontinuum and the amount of glass used in our setup, these modes are difficult to resolve. However, by tuning the probe spectrum and maximizing the amount of light near -1500 cm$^{-1}$, we are in fact able to resolve these weaker modes, as shown in Fig. 4(b). More specifically, we observe the following modes: -1440 cm$^{-1}$(CH$_2$ scissoring), -1296 cm$^{-1}$(CH$_2$ wagging), -1116 cm$^{-1}$(likely CH$_3$ rocking), and -1068 cm$^{-1}$ which could be the CC stretching, CO stretching, or CH$_2$ rocking mode. The OH stretching vibrations from -3600 to -3100 cm$^{-1}$ are not detected due to their reportedly low Raman cross-sections compared to the CH stretching modes.[39]

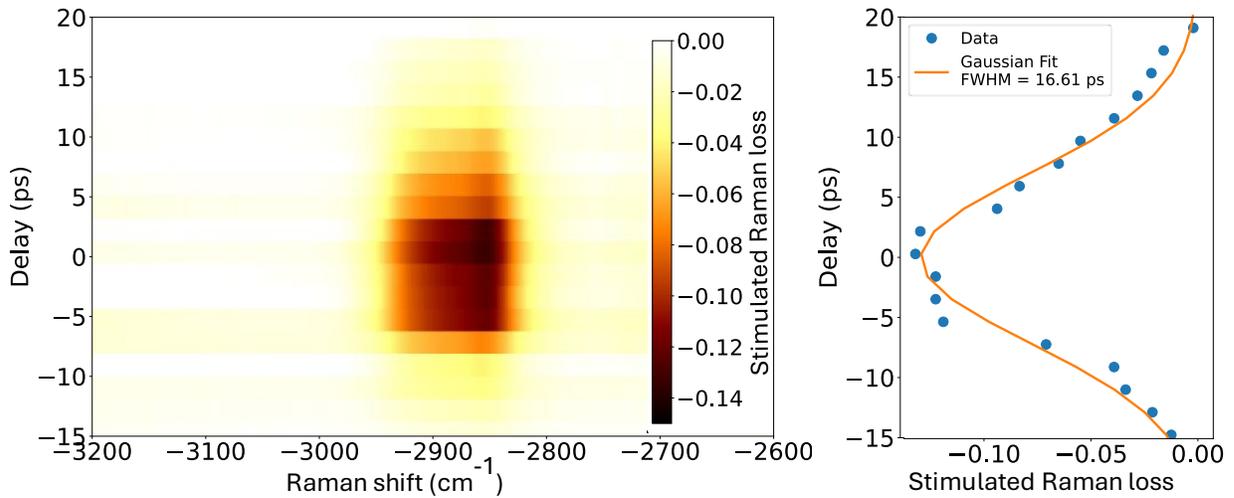

Figure 5: (a) Stimulated Raman loss as a function of pump delay (optical path difference) in picoseconds. (b) The blue points are a lineout of (a) at -2858 cm$^{-1}$. We fit a Gaussian function (orange line) to the data and find a full width at half-maximum (FWHM) of approximately 16.6 ps.

Fig. 5(a) shows the Raman loss signal as a function of delay between the Raman pump and probe. We convert the position of the delay stage in the pump arm to an optical path difference in picoseconds. In Fig. 5(b), we take a vertical slice of (a) at the centre of the Raman signal, approximately –2858 cm$^{-1}$, and fit a Gaussian function to the data, yielding a full width at half-maximum (FWHM) of approximately 16.6 ps.

The group velocity mismatch between the pump and probe may also contribute to this lifetime; however, based on our previous work, we calculate a mismatch of only 1.24 ps, which is small compared to the measured signal duration. This indicates that the signal is primarily determined by the temporal overlap of the pump and probe pulses. The 16.6 ps lifetime is slightly longer than the 6 ps pump pulse, suggesting an additional contribution from the Raman dephasing time that warrants further study.

**Conclusion**

We demonstrate a new method for generating a Raman probe pulse, Kerr-instability amplification for broadband stimulated Raman spectroscopy (KAB-SRS), which improves upon existing OPA-based approaches in tunability, versatility of gain media, and cost-efficiency. By directly amplifying a supercontinuum spectrum spanning 530–1000 nm (–6000 to 2800 cm$^{-1}$), KAB-SRS enables the use of a wide range of Raman pump wavelengths. Using a blue-shifted probe, we measure stimulated Raman loss signals and resolve the strongest vibrational modes of 1-decanol between –3000 and –1000 cm$^{-1}$.

We expect this method to provide a practical pathway for developing next-generation FSRS systems, and we focus future work on dispersion compensation to enable the simultaneous detection of multiple Raman signals across a broad spectrum.

**Acknowledgements**

We thank Aaron Fisk for useful conversations. We acknowledge the technical assistance of Pratik Choudhari. We thank Steven J. Rehse for advice on the preparation of this manuscript.

**Declaration of conflicting interest**

The authors declared no potential conflicts of interest with respect to the research, authorship, and/or publication of this article.

**Funding statement**

N.G. Drouillard acknowledges support from the Ontario Graduate Scholarship and TJ Hammond acknowledges funding from the Natural Sciences and Engineering Research Council of Canada (RGPIN-2019-06877) and the University of Windsor Xcellerate grant (5218522). The authors also acknowledge financial support from CRC-2023-0089.